\documentstyle[12pt,aasms4,psfig]{article}
\begin{document}
\title{THE COSMOLOGICAL MASS DISTRIBUTION FUNCTION IN THE ZEL'DOVICH
APPROXIMATION}
\author{Jounghun Lee and Sergei F. Shandarin} 
\affil{Department of Physics and Astronomy,\\
University of Kansas, Lawrence, KS 66045 \\
taiji, sergei@kusmos.phsx.ukans.edu}
\begin{abstract}

An analytic approximation to the mass function for gravitationally bound
objects is presented. We base on the Zel'dovich approximation 
to extend the Press-Schechter formalism to a nonspherical 
dynamical model.  A simple extrapolation of that approximation 
suggests that the gravitational collapse along all three directions    
which eventually leads to the formation of real virialized
objects - clumps occur in the regions where the lowest eigenvalue
of the deformation tensor, $\lambda_{3}$, is positive.
We derive the conditional probability of $\lambda_{3} > 0$ 
as a function of the linearly extrapolated density contrast 
$\delta$ and the conditional probability distribution of $\delta$  
provided that $\lambda_{3}>0$.
These two conditional probability distributions show that 
the most probable density of the bound regions ($\lambda_{3}>0$)
is roughly $1.5$ at the characteristic mass scale $M_{*}$, 
and that the probability of $\lambda_{3} > 0$ is almost unity 
in the highly overdense regions ($\delta > 3\sigma$).
Finally an analytic mass function of clumps is derived with a help
of one simple $ansatz$ which is employed to treat the multistream
regime beyond the validity of the Zel'dovich approximation.
The resulting mass function is renormalized by a factor 
of $12.5$, which we justify with a sharp k-space filter 
by means of the modified Jedamzik analysis.
Our mass function is shown to be different from the 
Press-Schechter one, having a lower peak and predicting more 
small-mass objects. 


\end{abstract}

\section{INTRODUCTION} 

Our present universe is observed to be quite clumpy with numerous galaxies,
groups of galaxies, galaxy clusters, and etc. which span a large
dynamic range in mass.
The mass distribution function of these large scale structures is the 
crucial key to the nature of primordial density fluctuations from which the
cosmic structures are believed to have arisen through gravitational 
growth, recollapse, and virialization (\cite{kol-tur90}).
Since these gravitational processes are inherently nonlinear 
and sufficiently complicated, it is not an easy task to find the mass 
distribution function for bound objects analytically.
Owing to its important role in cosmology, however, much effort has been
made on determining even an approximate expression of the mass function
(e.g., \cite{pee85}; \cite{wil-etal91}; \cite{bra-vil92}; \cite{cav-men94}; 
\cite{ver-etal94}; Cavaliere, Menci, \& Tozzi 1996). 
For recent review, see Monaco (1997).  

The pioneering attempt in this field has been ascribed 
to Press $\&$ Schechter 
(1974, hereafter PS) \footnote{See also Doroshkevich (1967).}
who proposed an analytic formalism for the mass function based on 
two simple assumptions: 1) the initial density field is Gaussian;
2) the gravitational collapse of mass elements is spherical and  
homogeneous.  
Along with these two assumptions, 
PS also postulated that the number densities
of bound objects could be counted by filtering the initial linear
density field.
Although much criticism thereafter was poured upon the PS formalism
about its unrealistic treatment of the collapse process and 
unclarified arguments including the $notorious$ normalization  
factor of 2,  the PS mass function has survived many numerical tests,  
showing good agreement with the results from N-body simulations 
(e.g., \cite{efs-etal88}; \cite{bon-etal91}; \cite{la-co94}). 

Motivated by somewhat unexpected success of the PS mass function, many
authors have tried to understand why it works so well in practice.
Peacock $\&$ Heaven (1990) and Bond et al. (1991) have shown, 
by using the excursion set theory, 
that the $fudge$ factor of 2 in the PS formalism  which is 
directly related to the {\it cloud-in-cloud} problem can be justified 
with a sharp k-space filter. Jedamzik (1995) solved this cloud-in-cloud 
problem by means of the integral equation for the  mass function.  He  
insisted that the PS mass function should be altered even in the case of 
a sharp k-space filter. 
Yet Yano, Nagashima, $\&$ Gouda (1996) have argued that the sharp 
k-space filter recovers the PS mass function with the normalization
factor of 2 even in the Jedamzik formalism 
if a mathematically consistent definition of
isolated bound objects is used and the spatial correlations are neglected.
They have also shown by introducing the two point correlation function 
into the Jedamzik formalism that the possible overlapping effect of 
density fluctuations which is responsible for the  
fragmentation and the coagulation of bound objects 
(see \cite{sil-wh78}; \cite{lu88}; \cite{cav-men93}) 
can be neglected either on very small or on large mass scales.  

The PS approach has been also applied to nonspherical dynamical 
models.  Monaco (1995) has suggested that the 
mass function should be treated as a Lagrangian quantity.
Employing the Zel'dovich approximation as a proper Lagrangian
dynamics, he computed the collapse epoch along the first
principal axis, and showed that the shear shortens the collapse
time and thus more high-mass structures are expected to form
than the original PS mass function predicts.  This effect of
the shear explains dynamically the lowered density threshold
$(\delta_{c} \simeq 1.5)$ detected in several N-body experiments 
(e.g., \cite{efs-ree88}; \cite{car-couch89}; 
\cite{kly-etal95}; \cite{bo-my96}).    

Shandarin $\&$ Klypin (1984) have shown by N-body simulations that 
nonlinear clumps form from the Lagrangian regions where 
the smallest eigenvalue of the deformation tensor, $\lambda_{3}$,  
reaches a local maximum.
Recently, Audit, Teyssier, $\&$ Alimi (1997) have 
proposed some analytic prescriptions to compute
the collapse time along the second and the third principal
axes, pointing out that Lagrangian dynamics is not valid 
after the first axis collapse but the formation of real virialized 
clumps must correspond to the third axis collapse.  
Their argument agrees with the N-body result
obtained by Shandarin $\&$ Klypin (1984).  
In their analysis, the shear delays the third axis collapse rather
than fastens it in contrast to its effect on the first axis collapse, 
which is in agreement with Peebles (1990).  

The normalization problem, however, has not been well addressed 
in these nonspherical approaches to the mass function.  
Monaco (1995) adopted the normalization factor 
of 2 used in the PS formalism, 
while Audit et al. (1997) just assumed that the mass
function could be normalized properly in any case.

In this paper we study the eventual formation of clumps 
in a spatially flat matter-dominated universe, 
with fragmentation and coagulation effects ignored. 
In $\S$ 2 we review the statistical treatment of the mass function, 
highlighting the PS formalism.
In $\S$ 3 a nonspherical approach to the collapse condition 
based on the Zel'dovich approximation is described, and two 
useful conditional probability distributions relating the 
density field to the collapse condition are derived. In $\S$ 4 
an $ansatz$ is proposed to extend the validity of the 
Lagrangian dynamics to the third axis collapse.  
With a help of this ansatz 
an analytic approximation to the mass function for clumps 
is derived.  In $\S$ 5 we justify the normalization factor $12.5$ of 
the resulting mass function by using the Jedamzik integral equation.  
In $\S$ 6 the results are discussed and final conclusions  
are drawn. 
We relegate the detailed calculations 
and derivations to two Appendices.

\section{STATISTICAL DESCRIPTION OF MASS FUNCTIONS}

The mass function $n(M)$ is defined such that $n(M)dM$ is the comoving 
number density of gravitationally bound objects 
in the mass range $(M,M + dM)$.
To compute this statistics, 
it is assumed that the number densities of bound
objects can be inferred from the linearly extrapolated density
contrast field, $\delta \equiv \delta\rho/\bar{\rho}$ 
($\bar{\rho}$: mean density).
In other words, if a given region of the linear density field  
satisfies a specified criterion of collapse, then it is supposed 
to collapse and form a bound object.

Let $F(M)$ be the probability of finding a region satisfying a given 
collapse condition in the linear  
density field filtered at mass scale $M$, or 
equivalently the fraction of the volume occupied by the regions which 
will eventually collapse into bound objects with masses greater than $M$.
Then we may write $F(M)$ as follows:
\begin{equation}
F(M) = \int^{\infty}_{-\infty}\! p(\delta)\cdot C d\delta.
\end{equation}
Here $p(\delta)d\delta$ is the probability
that the smoothed density field at any given point
will have a value in the range $(\delta, \delta + d\delta)$, and
$C$ stands for the probability that the chosen point with
density $\delta$ will actually collapse.
Once $p(\delta)$ and $C$ are determined, and then $F(M)$ is found, 
the mass function $n(M)$ can be easily obtained as
\begin{equation}
n(M) = \frac{\bar{\rho}}{M}\bigg{|}
\frac{dF}{dM}\bigg{|} ,
\end{equation}
where $M/\bar{\rho}$ is nothing but the volume of a bound
region with mass $M$.

The specific functional form of $C$ is determined by the chosen
dynamics to explain the collapse process, while $p(\delta)$
depends on the property of the initial density field which
is often assumed to be Gaussian in the standard cosmology
(see \cite{bar86}).
The probability distribution of the Gaussian density field
smoothed out by a window function $W(R)$ of scale radius
$R$ is given by 
\begin{equation}
p(\delta) = \frac{1}{\sqrt{2\pi}\sigma(M)}
\exp\bigg{[}-\frac{\delta^2}{2\sigma^2(M)}\bigg{]}.
\end{equation}
Here the mass variance $\sigma^{2}(M)$ is a function of 
scale mass $M \propto \bar{\rho}R^3$ and estimated by
\begin{equation}
\sigma^{2}(M) = \int\! \frac{d^3k}{(2\pi)^3}
|\delta_{k}|^2 W_{k}^{2}(R) ,
\end{equation}
where $\delta_{k}$ and $W_{k}(R)$ are the Fourier components of the
density $\delta$ and the window function $W(R)$ respectively.

According to the top-hat spherical model adopted by PS,
the bound objects form in the regions where the linearly
extrapolated density contrast $\delta$, growing with time, 
reaches its critical value $\delta_{c} \simeq 1.69$ 
in a flat universe (\cite{pee93}).
Therefore the regions with $\delta > \delta_{c}$ when 
filtered at scale radius $R$ correspond to the bound
objects with masses greater than $M(R)$ since
it will have $\delta = \delta_{c}$ when filtered at 
some larger scale.
Thus in the PS formalism the collapse probability $C$
in equation (1) is determined solely by the density field 
itself, and can be expressed by the following Heavyside step function:
\begin{equation}
C_{ps} = \Theta(\delta - \delta_{c}).
\end{equation}
Using equations (1), (3), and (5), one obtains
\begin{eqnarray}
F(M) &=& \frac{1}{\sqrt{2\pi}\sigma(M)}\int^{\infty}_{-1}\!
\exp\bigg{[}-\frac{\delta^2}{2\sigma^2(M)}\bigg{]}
\Theta(\delta - \delta_{c})d\delta
\nonumber \\
&=& \frac{1}{\sqrt{2\pi}\sigma(M)}\int^{\infty}_{\delta_{c}}\!
\exp\bigg{[}-\frac{\delta^2}{2\sigma^2(M)}\bigg{]}d\delta \nonumber \\
&=& \frac{1}{2}{\rm erfc}\bigg{[}\frac{\delta_{c}}{\sqrt{2}\sigma(M)}
\bigg{]},
\end{eqnarray}
where erfc(x) is the complementary error function.

One obvious problem with the above analysis is that the integral
of $dF/dM$ 
over the whole range of mass does not give unity:
\begin{equation} 
\int^{\infty}_{0}\!\frac{dF}{dM}dM 
= \int^{\infty}_{0}dF = \frac{1}{2}.
\end{equation}

This normalization problem is originated from the fact that
the PS formalism does not account for the underdense regions
properly. Even for regions with $\delta < \delta_{c}$ at a given 
filtering scale, there is still a $nonzero$ probability
that such regions will have $\delta > \delta_{c}$ when filtered
at some larger scale. But the PS formalism completely ignored
those underdense regions in estimating $F(M)$, so half the
mass initially present in the underdense regions was not taken
care of.  
PS avoided this normalization problem simply by multiplying 
$dF/dM$ by a factor of $2$, and wrote 
the mass function in the form such that   
\begin{eqnarray}
n_{ps}(M) &=& 2\frac{\bar{\rho}}{M}\Bigg{|}
\frac{dF}{dM}\Bigg{|} 
 = 2\frac{\bar{\rho}}{M}\Bigg{|}\frac{d\sigma}{dM}
\frac{\partial F}{\partial \sigma}\Bigg{|} \nonumber \\
&=& \sqrt{\frac{2}{\pi}}
\frac{\bar{\rho}}{M}\Bigg{|}\frac{d\sigma}{dM}\Bigg{|}
\frac{\delta_{c}}{\sigma^2(M)}\exp\bigg{[}-\frac{\delta_{c}^2}
{2\sigma^2(M)}\bigg{]}.
\end{eqnarray}

As mentioned in $\S$ 1, the $cooked$ $up$ 
normalization factor of $2$ in equation (8) has been shown to be 
correct in the case of a sharp k-space filter  
[$W_{k}(R) = \Theta(\pi/R-k)$], 
and various numerical tests have confirmed the PS mass function
as a satisfactory approximation. 
Nevertheless it still leaves much to be desired: the physical meaning
of the sharp k-space filter has yet to be understood; the 
gravitational collapse should be treated in more realistic models 
than the top-hat spherical one; 
the lowered density threshold $(\delta_{c} \simeq 1.5)$
obtained in many numerical tests 
cannot be explained by this statistical argument, and so the PS  
mass function is degraded to a phenomenological device.

\section{NONSPHERICAL APPROACH TO COLLAPSE CONDITION}

Since PS derived their mass function on the basis
of the top-hat spherical model in 1974, the nonspherical nature 
of the gravitational collapse has been demonstrated by many authors 
(e.g., \cite{sh-etal95}; Kuhlman, Melott, $\&$ Shandarin 1996).
Especially the shear has been shown to 
play a very important role in the formation of the nonlinear 
structures 
(e.g., \cite{pee90}; \cite{mo95}; \cite{au-etal97}).
Therefore it is necessary to consider more realistic
dynamical models to understand the collapse process 
and find the mass function.

We choose the Zel'dovich approximation as a suitable 
Lagrangian dynamics to take into account the nonspherical aspect of 
the gravitational collapse. However, 
instead of bringing the effect of the shear up to the surface, 
we try to retain the framework of the PS formalism, counting 
the number densities of bound objects from the filtered linear   
density field but with a different dynamical collapse probability 
$C$ in equation (1). 

\subsection{The Zel'dovich Approximation}

The Zel'dovich approximation (Zel'dovich 1970) asserts that 
the trajectory of a cosmic particle in the comoving coordinates
can be expressed by the following simple formula:
\begin{equation}
{\bf x} = {\bf q} - D_{+}(t)\nabla \Psi ({\bf q}).
\end{equation}
Here ${\bf q}$ and  ${\bf x}$ are the Lagrangian (initial) and the Eulerian 
(final) coordinates of the particle respectively, 
$\Psi ({\bf q})$
is the perturbation potential which is a Gaussian random field, and 
$D_{+}(t)$ describes the growth of density fluctuations as a function of 
time.  Throughout this paper, we focus on a spatially flat 
matter-dominated universe with vanishing cosmological constant,  
in which case $D_{+}(t) \propto a(t) \propto t^{2/3}$ [$a(t)$: the 
cosmic expansion factor].   

Applying a simple mass conservation relation $\bar{\rho}d^{3}{\bf q} = 
\rho({\bf x})d^{3}{\bf x}$ to the above formula (9) gives the 
following expression of the mass density:
\begin{equation}
\rho({\bf x}) = \frac{\bar{\rho}}
{[1 - D_{+}(t)\lambda_{1}({\bf q})][1 - D_{+}(t)
\lambda_{2}({\bf q})][1 - D_{+}(t)\lambda_{3}({\bf q})]} ,
\end{equation}
where $\lambda_{1}$, $\lambda_{2}$, $\lambda_{3}$ are the 
ordered eigenvalues ($\lambda_{1}>\lambda_{2}>\lambda_{3}$)
of the deformation tensor, 
\begin{equation}
d_{ij} = \frac{\partial^{2}\Psi}{\partial q_{i}\partial q_{j}}.
\end{equation}
Equation (10) shows that three random fields 
$\lambda_{1}({\bf q}),\lambda_{2}({\bf q}),\lambda_{3}({\bf q})$
in the Lagrangian space 
are now the new dynamic quantities determining the collapse condition
of given cosmic masses in the corresponding Eulerian space.
Thus the mass function of bound objects can be built upon this
Lagrangian dynamical theory (see also \cite{mo95}).

The actual dynamics for the formation of gravitationally 
bound objects is very complex. 
Even in the frame of the Zel'dovich approximation, the description 
of the gravitational collapse along all three directions is 
far from being simple and too cumbersome to use 
(Arnol'd, Shandarin, $\&$ Zel'dovich 1982). 
Here we employ rather a simplified dynamical model to 
approximate the collapse process and determine the collapse 
condition for the formation of clumps.
 
Provided that at least one of the eigenvalues is positive 
at a given (Lagrangian) point, the denominator in equation (10) 
can become zero as $D_{+}(t)$ increases with time, so the density 
$\rho({\bf x})$ will diverge, signaling collapse at the corresponding 
Eulerian point. 
If only the largest eigenvalue is positive ($\lambda_{1}>0, 
\lambda_{3}<\lambda_{2}<0$) in a given region, 
then it collapses into a pancake.
If two eigenvalues are positive ($\lambda_{1}>\lambda_{2}>0$) while  
the third one is negative ($\lambda_{3}<0$), then a filament forms.
The formation of a virialized bound object -- a clump occurs  
only if all of three eigenvalues are positive, i.e.  
$\lambda_{3}>0$. 
So, in our dynamical model based on the Zel'dovich approximation, 
it is assumed that the lowest eigenvalue, $\lambda_{3}$, plays the most 
crucial role in determining the collapse condition for the formation 
of clumps.  This assumption is in general agreement with 
Shandarin $\&$ Klypin (1984). 

The useful joint probability distribution of an ordered set 
($\lambda_{1},\lambda_{2},\lambda_{3}$) is derived by Doroshkevich (1970): 
\begin{equation}
p(\lambda_{1},\lambda_{2},\lambda_{3}) = 
\frac{3375}{8\sqrt{5}\pi\sigma^6}\exp\bigg{(}-\frac{3I_{1}^2}{\sigma^2}
 + \frac{15I_{2}}{2\sigma^2}\bigg{)}(\lambda_{1}-\lambda_{2})
(\lambda_{2}-\lambda_{3})(\lambda_{1}-\lambda_{3}) ,
\end{equation}
where $I_{1} = \lambda_{1}+\lambda_{2}+\lambda_{3}$, $I_{2} = 
\lambda_{1}\lambda_{2} + \lambda_{2}\lambda_{3} + \lambda_{3}\lambda_{1}$, 
and $\sigma^2$ is the mass variance as defined 
in equation (4). From equation (12), one can see that the 
Zel'dovich approximation excludes both exactly spherical   
($\lambda_{1}=\lambda_{2}=\lambda_{3}$)
and exactly cylindrical  
($\lambda_{1}=\lambda_{2}$, $\lambda_{2}=\lambda_{3}$,  
$\lambda_{3}=\lambda_{1}$) collapse. 
Both types of collapse have zero probability of occurring.   
(However, the points with $\lambda_{i}=\lambda_{j}$ 
exist in generic fields on lines, 
that is on a set of measure zero in 3-dim., while the points with 
$\lambda_{1}=\lambda_{2}=\lambda_{3}$ do not exist at all.) 

In order to obtain deeper qualitative understanding of the collapse 
in the Zel'dovich approximation, it may be also useful, in addition to 
this joint probability distribution (12), to have 
individual probability distribution of each eigenvalue
\footnote{Doroshkevich (1970) derived the probability distribution of 
$\lambda_{1}$.  But  we found out a typo in  his result. 
Except for the typo, equation (13) agrees with his result.}
(see Appendix A):
\begin{eqnarray}
p(\lambda_{1}) & = &
\frac{\sqrt{5}}{12\pi\sigma}\Bigg{\{} 20\frac{\lambda_1}{\sigma}
\exp\bigg{(}-\frac{9\lambda_{1}^2}{2\sigma^2}\bigg{)}
- \sqrt{2\pi}\exp\bigg{(}-\frac{5\lambda_{1}^2}{2\sigma^2}\bigg{)}
{\rm erf}\bigg{(}\sqrt{2}\frac{\lambda_{1}}{\sigma}\bigg{)}\bigg{(}1-20
\frac{\lambda_{1}
^2}{\sigma^2}
\bigg{)} \nonumber \\
&&- \sqrt{2\pi}\exp\bigg{(}
-\frac{5\lambda_{1}^2}{2\sigma^2}\bigg{)}\bigg{(}1-20
\frac{\lambda_{1}
^2}{\sigma^2}\bigg{)}
+3\sqrt{3\pi}\exp\bigg{(}-\frac{15\lambda_{1}^2}{4\sigma^2}\bigg{)}
{\rm erf}\bigg{(}\frac{\sqrt{3}\lambda_{1}}{2\sigma}\bigg{)} \nonumber  \\
&&
+ 3\sqrt{3\pi}\exp\bigg{(}-
\frac{15\lambda_{1}^2}{4\sigma^2}\bigg{)}\Bigg{\}}, \\
\nonumber \\
p(\lambda_{2}) & = & 
\frac{\sqrt{15}}{2\sqrt{\pi}\sigma}\exp\bigg{(}-\frac{15\lambda_{2}^2}
{4\sigma^2}\bigg{)}, \\
\nonumber \\
p(\lambda_{3}) & = &
-\frac{\sqrt{5}}{12\pi\sigma}\Bigg{\{} 20\frac{\lambda_3}{\sigma}
\exp\bigg{(}-\frac{9\lambda_{3}^2}{2\sigma^2}\bigg{)}
+ \sqrt{2\pi}\exp\bigg{(}-\frac{5\lambda_{3}^2}{2\sigma^2}\bigg{)}
{\rm erfc}\bigg{(}\sqrt{2}\frac{\lambda_{3}}{\sigma}\bigg{)}\bigg{(}1-20
\frac{\lambda_{3}^2}{\sigma^2}\bigg{)} \nonumber \\
&&- 3\sqrt{3\pi}\exp\bigg{(}-\frac{15\lambda_{3}^2}{4\sigma^2}\bigg{)}
{\rm erfc}\bigg{(}\frac{\sqrt{3}\lambda_{3}}{2\sigma}\bigg{)}\Bigg{\}}. 
\end{eqnarray}
The above individual
probability distributions (13), (14), and (15) for the rescaled 
variable $\lambda/\sigma$ are plotted in Figure 1.  Note that 
the distribution of $\lambda_{2}({\bf q})$ is Gaussian 
despite that $\lambda_{2}({\bf q})$ is not a Gaussian random field.

According to equation (15), $\lambda_{3} > 0$ has a low probability 
of occurring, 0.08 (see \cite{do70}, or Appendix A).  
However, the small value of $P(\lambda_{3} > 0) = 0.08$
does not indicate that only $8 \% $ of the whole 
regions will collapse into clumps. But rather it indicates that 
the probability of finding a bound region  
at filtering mass scale $M$ is 0.08, provided that it is 
included in an isolated bound object with larger mass 
$M^{\prime} > M$ (see $\S$ 5). 
Here the isolated bound objects indicate the bound objects  
which have just collapsed at a given epoch.  

In the following subsection, we derive the conditional probabilities 
of $\lambda_{3}>0$ and $\delta$, reveal the correlated properties 
between them, and determine a nonspherical collapse probability $C$.  

\subsection{Conditional Probabilities}

In the linear regime when $D_{+}(t)$ is still less than unity, 
equation (10) can be approximated by
\begin{equation}
\rho \simeq\bar{\rho}[1 + D_{+}(\lambda_{1} + \lambda_{2} + \lambda_{3})].
\end{equation}
Setting $D_{+}\equiv 1$ at the present epoch, the linearly
extrapolated density contrast is now written as 
\begin{equation}
\delta  = \frac{\delta\rho}{\bar{\rho}} = 
\lambda_{1} + \lambda_{2} + \lambda_{3}.
\end{equation}
Let us choose $(\delta,\lambda_{2},\lambda_{3})$ as a new set of variables.
Then equation (12) can be reexpressed as a joint probability distribution 
of $(\delta,\lambda_{2},\lambda_{3})$ such that
\begin{eqnarray}
p(\delta,\lambda_{2},\lambda_{3}) & = & \frac{3375}{8\pi\sqrt{5}\sigma^6}
\exp\bigg{[}-\frac{3\delta^2}{\sigma^2} + \frac{15}{2\sigma^2}
(\lambda_{2}+\lambda_{3})(\delta-\lambda_{2}-\lambda_{3})
 + \frac{15}{2\sigma^2}\lambda_{2}\lambda_{3}\bigg{]} \nonumber \\
&&\times
(\delta - 2\lambda_{2} - \lambda_{3})(\lambda_{2} - \lambda_{3})
(\delta - \lambda_{2} - 2\lambda_{3}).
\end{eqnarray}
Direct integration of the above joint distribution (18) over $\lambda_{2}$
gives the two point probability distribution of ($\delta$, 
$\lambda_{3}$):
\begin{eqnarray}
p(\delta,\lambda_{3}) &=& \int^{\frac{\delta-\lambda_{3}}{2}}_{\lambda_{3}}
p(\delta,\lambda_{2},\lambda_{3})d\lambda_{2} \nonumber \\
&=&\frac{3\sqrt{5}}{16\pi\sigma^4}\bigg{(}15\delta^2-90\lambda_{3}\delta
+135\lambda_{3}^2-8\sigma^2\bigg{)}\exp\bigg{(}-\frac{9\delta^2
-30\lambda_{3}\delta+45\lambda_{3}^2}{8\sigma^2}\bigg{)} \nonumber \\
&&  +  \frac{3\sqrt{5}}{2\sigma^2\pi}\exp
\bigg{(}-\frac{6\delta^2-30\lambda_{3}\delta+45\lambda_{3}^2}{2\sigma^2}
\bigg{)},
\end{eqnarray}
where the upper limit and the lower limit of $\lambda_{2}$ are
$(\delta-\lambda_{3})/2$ and $\lambda_{3}$ respectively due to the  
condition of $\lambda_{1}>\lambda_{2}>\lambda_{3}$.

With equation (19), we can investigate various correlated properties
between $\delta$-field and $\lambda_{3}$-field. First of all, let
us calculate the probability distribution of $\delta$ confined
in the regions with $\lambda_{3}>0$: 
\begin{eqnarray}
p(\delta|\lambda_{3}>0) &=& \frac{p(\delta,\lambda_{3}>0)}{P(\lambda_{3}>0)}
 = \frac{\int\!^{\frac{\delta}{3}}_{0}
p(\delta,\lambda_{3})d\lambda_{3}}
{P(\lambda_{3}>0)} \nonumber \\
&=& \bigg{\{}-\frac{75\sqrt{5}}{8\pi\sigma^2}\delta\exp\Big{(}
-\frac{9\delta^2}{8\sigma^2}\Big{)}  \nonumber \\
&& +
\frac{25}{4\sqrt{2\pi}\sigma}\exp\Big{(}-\frac{\delta^2}{2\sigma^2}
\Big{)} 
\bigg{[}\/{\rm erf}\Big{(}\frac{\delta\sqrt{10}}{4\sigma}\Big{)}+
\/{\rm erf}\Big{(}\frac{\delta\sqrt{10}}{2\sigma}\Big{)}\bigg{]}
\bigg{\}}\Theta(\delta).
\end{eqnarray}
Here $\Theta$ stands for the Heavyside step function, and 
the condition $\lambda_{1}>\lambda_{2}>\lambda_{3}$ is used 
again to determine $\delta/3$ for the upper limit of $\lambda_{3}$.  
Figure 2 compares the unconditional Gaussian distribution of
the density field (3) with this conditional 
probability distribution (20) for the rescaled variable 
$\delta/\sigma$.  
It is shown that the maximum of $p(\delta|\lambda_{3}>0)$ is 
reached when $\delta \simeq 1.5\sigma$.
That is, the linearly extrapolated density of the regions 
satisfying $\lambda_{3}>0$
is most likely to be around $1.5\sigma$.
The average density contrast, $<\delta>_{\lambda_{3}>0}$, 
can be also computed with equation (20):
\begin{equation}
<\delta>_{\lambda_{3}>0} = \int^{\infty}_{0}\delta
p(\delta|\lambda_{3}>0)d\delta
= \frac{25\sqrt{10}\sigma}{144\sqrt{\pi}}(3\sqrt{6}-2)
\simeq 1.65\sigma .
\end{equation}
So in the regions with $\lambda_{3} > 0$,
the average density $<\delta>_{\lambda_{3}>0}$ is slightly higher
than the most probable density, say
$\delta_{\lambda_{3}>0}^{max}$.  We note that for $\sigma = 1$,  
$\delta_{\lambda_{3}>0}^{max}$ roughly coincides with 
the lowered density threshold $\delta_{c}\simeq 1.5$ of the
PS mass function, while
$<\delta>_{\lambda_{3}>0}$ is close to the spherical
threshold value $\delta_{c}\simeq 1.69$.
Setting $\sigma=1$ means filtering the density 
field on characteristic mass scale $M_{*}$
[defined by $\sigma(M_{*})=1$].  Thus the regions with $\lambda_{3}>0$
for $\sigma=1$ correspond to clumps with masses $M > M_{*}$.
In fact, as argued by Monaco (1995),  it is unavoidable to 
limit our Lagrangian dynamical approach to the high-mass 
section $(M > M_{*})$ 
since the Zel'dovich approximation is valid only in the single 
stream regions, while the multistream regions are rare for 
$M > M_{*}$ (\cite{kof-etal94}).  

Another conditional distribution 
worth deriving is $P(\lambda_{3}>0|\delta)$, the  probability
that a given region with density $\delta$ will have all positive 
eigenvalues:
\begin{eqnarray}
P(\lambda_{3}>0|\delta) &=& \frac{p(\delta,\lambda_{3}>0)}{p(\delta)}
\nonumber \\
&=& \bigg{\{}-\frac{3\sqrt{10}}{4\sqrt{\pi}\sigma}\delta\exp\Big{(}
-\frac{5\delta^2}{8\sigma^2}\Big{)} + \frac{1}{2}
\bigg{[}{\rm erf}\Big{(}\frac{\delta\sqrt{10}}{4\sigma}\Big{)} 
+ {\rm erf}\Big{(}\frac{\delta\sqrt{10}}{2\sigma}\Big{)}\bigg{]}
\bigg{\}}\Theta(\delta) .
\end{eqnarray}
The resulting conditional probability (22) for the rescaled variable 
$\delta/\sigma$ is plotted in Figure 3. 
The probability of $\lambda_{3} >0$ 
begins to exceed one-half when $\delta \simeq 1.5\sigma$, and 
reaches unity when $\delta \simeq 3\sigma$. 
This implies that the collapse 
of highly overdense regions $(\delta \gg \sigma)$ will be 
always along all three directions (see also \cite{ber94}). 

We take equation (22) as our nonspherical collapse probability $C$ and 
proceed to derive the mass function of clumps analytically
in the next section.

\section{AN ANALYTIC APPROXIMATION TO MASS FUNCTIONS}

As noted earlier, the Zel'dovich approximation as a first order 
Lagrangian theory works very well till 
the first orbit crossing (corresponding to the formation of pancakes)
but breaks down afterwards in the multistream regime  
(\cite{sh-ze89}).  
Therefore the rather restrictive collapse condition purely 
based on this Lagrangian formalism may not be fully satisfactory
to describe the formation of clumps, especially low-mass objects.

On the other hand,  Shandarin $\&$ Klypin (1984) have shown by 
N-body simulations that the clumps  form from the Lagrangian regions 
where the smallest eigenvalue $\lambda_{3}$ of the deformation 
tensor reaches a local maximum.  
Thus, one practical way to overcome the limited validity of the Zel'dovich 
approximation within the framework of our dynamical approach
to mass functions is to parameterize the collapse 
condition by $\lambda_{3} > \lambda_{3c}$,  
assuming that the critical value of $\lambda_{3c}$
is a free parameter.  
Employing this simple $ansatz$ to derive $n(M)$, 
we first calculate 
the following probability distribution with equations (3) and (19) 
\begin{eqnarray}
P(\lambda_{3}>\lambda_{3c}|\delta) 
&=& \frac{p(\delta,\lambda_{3}>\lambda_{3c})}{p(\delta)}
= \frac{\int^{\frac{\delta}{3}}_{\lambda_{3c}}
p(\delta,\lambda_{3})d\lambda_{3}}{p(\delta)} \nonumber \\
&=& \Bigg{\{}-\frac{3\sqrt{10}}{4\sqrt{\pi}\sigma}(\delta-3\lambda_{3c})
\exp\bigg{[}-\frac{5(\delta-3\lambda_{3c})^2}{8\sigma^2}\bigg{]} \nonumber\\ 
&& + \frac{1}{2}\bigg{\{}{\rm erf}\bigg{[}
\frac{(\delta-3\lambda_{3c})\sqrt{10}}{4\sigma}\bigg{]}
+ {\rm erf}\bigg{[}\frac{(\delta-3\lambda_{3c})\sqrt{10}}
{2\sigma}\bigg{]}\bigg{\}}\Bigg{\}}\Theta(\delta-3\lambda_{3c}).
\end{eqnarray}
Comparison of equation (23) with equation (22) reveals that 
$P(\lambda_{3}>\lambda_{3c}|\delta)$ is just  
horizontally shifted along $\delta$-axis by $3\lambda_{3c}$ from 
$P(\lambda_{3}>0|\delta)$ with its shape unchanged. 

Consequently, this ansatz is mathematically
equivalent to parallel transformation of the density field itself
by $-3\lambda_{3c}$. Thus equation (1) 
is now expressed as follows:
\begin{eqnarray}
F(M) &=& \int^{\infty}_{-\infty}\!p(\delta+3\lambda_{3c})\cdot
P(\lambda_{3}>0|\delta)d\delta
\nonumber \\
&=& \frac{1}{\sqrt{2\pi}\sigma}\int^{\infty}_{0}
\exp\Big{[}-\frac{(\delta+3\lambda_{3c})^2}{2\sigma^2}\Big{]}
\bigg{\{}-\frac{75\sqrt{10}}{8\sqrt{\pi}\sigma}\delta\exp\Big{(}
-\frac{5\delta^2}{8\sigma^2}\Big{)}
\nonumber \\
&& + \frac{25}{4}\bigg{[}
{\rm erf}\Big{(}\frac{\delta\sqrt{10}}{4\sigma}\Big{)} + 
{\rm erf}\Big{(}\frac{\delta\sqrt{10}}{2\sigma}\Big{)}\bigg{]}
\bigg{\}}d\delta .
\end{eqnarray}
Here the volume fraction $F(M)$ is normalized by a factor of $1/0.08=12.5$ 
which we justify with a sharp k-space filter in $\S$ 5.  
This normalization factor is much larger   
than the factor of 2 in the PS formalism.  
However, this larger normalization factor can be explained by the  
larger amount of {\it cloud-in-cloud} occurrences in our 
dynamical formalism than in the PS formalism,  
as shown in $\S$ 5 where such amount is computed.
In an ideal hierarchical model, all the masses are included in clumps.  
According to our dynamical model, 
only about 8\% of all the masses are included in the clumps 
with the "largest" mass (the ``largest mass''
of bound objects in the universe is, in a practical sense, 
$M \simeq M_{*}$).   
This is in rough agreement with the fraction of the galaxies in the 
Abell clusters (e.g., \cite{pad93}). 
All the remaining masses are included in the clumps at smaller filtering 
mass scales.  

Differentiating equation (24) with respect to $\sigma$, we have
\begin{eqnarray}
\frac{\partial{F}}{\partial{\sigma}} &=& 
\frac{\partial}{\partial\sigma}\Bigg{\{}
\frac{1}{\sqrt{2\pi}\sigma}\int^{\infty}_{0}
\exp\Big{(}-\frac{(\delta+3\lambda_{3c})^2}{2\sigma^2}\Big{)}
\bigg{\{}-\frac{75\sqrt{10}}{8\sqrt{\pi}\sigma}\delta\exp\Big{(}
-\frac{5\delta^2}{8\sigma^2}\Big{)} \nonumber \\
&&\hspace{6cm}
 + \frac{25}{4}\bigg{[}{\rm erf}\Big{(}\frac{\delta\sqrt{10}}{4\sigma}
\Big{)} + {\rm erf}\Big{(}\frac{\delta\sqrt{10}}{2\sigma}\Big{)}\bigg{]}
\bigg{\}}d\delta\Bigg{\}},  \nonumber \\
&=&\frac{25\sqrt{10}\lambda_{3c}}{2\sqrt{\pi}\sigma^2}
\Big{(}\frac{5\lambda_{3c}^2}{3\sigma^2}-\frac{1}{12}\Big{)}
\exp\Big{(}-\frac{5\lambda_{3c}^2}{2\sigma^2}\Big{)}
{\rm erfc}\Big{(}\frac{\sqrt{2}\lambda_{3c}}{\sigma}\Big{)}
\nonumber \\
&& +\frac{25\sqrt{15}\lambda_{3c}}{8\sqrt{\pi}\sigma^2}
\exp\Big{(}-\frac{15\lambda_{3c}^2}{4\sigma^2}\Big{)}
{\rm erfc}\Big{(}\frac{\sqrt{3}\lambda_{3c}}{2\sigma}\Big{)}
-\frac{125\sqrt{5}\lambda_{3c}^2}{6\pi \sigma^3}
\exp\Big{(}-\frac{9\lambda_{3c}^2}{2\sigma^2}\Big{)} . 
\end{eqnarray}
Figure 4 shows the generic behavior of this differential volume
fraction (25) as $\lambda_{3c}$ changes.
Since $\partial F/\partial \sigma$ is directly proportional to 
$n(M)$, one can conclude from Figure 4 that as $\lambda_{3c}$ 
increases, the number densities of small-mass clumps (large $\sigma$)
increase while the large masses (small $\sigma$) are reduced and 
the peak is lowered. 

For simple power law spectra $|\delta_{k}|^2 \propto k^n$,  
the mass variance becomes
\begin{equation}
\sigma^{2}(M) = \Bigg{(}\frac{M}{M_{*}}\Bigg{)}^{-(n+3)/3} .
\end{equation}
So in this case, the mass function can be expressed explicitly
in terms of $M$: 
\begin{eqnarray}
n(M) &=& \frac{\bar{\rho}}{M}\Bigg{|}\frac{dF}{dM}\Bigg{|}
= \frac{\bar{\rho}}{M}\Bigg{|}\frac{d\sigma}{dM}
\frac{\partial F}{\partial \sigma}\Bigg{|} \nonumber \\
&=&\frac{25\sqrt{10}\lambda_{3c}}{2\sqrt{\pi}}\bigg{(}\frac{n+3}{6}\bigg{)}
\frac{\bar{\rho}}{M^{2}}\bigg{(}\frac{M}{M_{*}}\bigg{)}^{(n+3)/6}
\times \nonumber \\
&&\Bigg{\{}
\bigg{[}\frac{5\lambda_{3c}^2}{3}
\bigg{(}\frac{M}{M_*}\bigg{)}^{(n+3)/3}-\frac{1}{12}
\bigg{]}\exp\bigg{[}-\frac{5\lambda_{3c}^2}{2}\bigg{(}\frac{M}{M_{*}}
\bigg{)}^{(n+3)/3}\bigg{]}
{\rm erfc}\bigg{[}\sqrt{2}\lambda_{3c}\bigg{(}\frac{M}{M_{*}}\bigg{)}
^{(n+3)/6}\bigg{]}
\nonumber \\
&&+ \frac{\sqrt{6}}{8}\exp\bigg{[}
-\frac{15\lambda_{3c}^2}{4}\bigg{(}\frac{M}{M_*}\bigg{)}^{(n+3)/3}
\bigg{]}{\rm erfc}\bigg{[}\frac{\sqrt{3}\lambda_{3c}}{2}
\bigg{(}\frac{M}{M_*}\bigg{)}^{(n+3)/6}\bigg{]}
\nonumber \\
&&-\frac{5\lambda_{3c}}{3\sqrt{2\pi}}
\bigg{(}\frac{M}{M_*}\bigg{)}^{(n+3)/6}\exp\bigg{[}
-\frac{9\lambda_{3c}^2}{2}\bigg{(}\frac{M}{M_*}\bigg{)}
^{(n+3)/3}\bigg{]}\Bigg{\}}.
\end{eqnarray}

We display the resulting mass function for 
$\lambda_{3c} \simeq 0.37$ in Figure 5.  The value of $0.37$ for 
$\lambda_{3c}$ is chosen to make our results for the high-mass section 
fit well with the $\delta_{c}\simeq 1.5$ PS mass function 
which has been  tested to be 
a good approximation (see \cite{mo95}).  The original PS mass
function with $\delta_{c} \simeq 1.69$ is also shown for comparison.
For every power index $n$ from $-2$ to $1$, the mass function (27) is
characterized by the following properties:

(1) In the high-mass section $(M/M_{*} > 1)$, it fits quite well with
the $\delta_{c} \simeq 1.5$ PS mass function.

(2) Its peak is lower than that of the PS one, which agrees with
N-body results (e.g., see \cite{efs-etal88}).
 
(3) It has approximately the same slope as the PS mass function but
predicts more structures in the low-mass section $(M/M_{*} < 1)$.

\section{NORMALIZATION}

Up to now, following the PS-like approach, we assumed that equation (2) 
is correct.  In other words, the probability of finding a region with
$\lambda_{3} > \lambda_{3c}$ at filtering mass scale $M$ is assumed
to be proportional to the fraction of the volume occupied by the
regions which will eventually collapse into bound objects with masses
$\ge M$.

Strictly speaking, however, equation (2) is not quite correct since  
the resulting mass function has to be always renormalized.  Even 
in the regions with $\lambda_{3} < \lambda_{3c}$ at the filtering mass
scale $M$, there is still a nonzero probability of $\lambda_{3} =
\lambda_{3c}$ when the density field is filtered at some larger scale
$M^\prime (> M$).  But this marginal probability is ignored by the
PS-like approach, which has resulted in a large normalization factor
$12.5$ of our mass function. 

Jedamzik (1995) suggested a generalization of equation (2) 
\footnote{In the original analysis based on the top-hat spherical 
model, Jedamzik (1995) did not use a 
mathematically correct definition of isolated bound objects.}. 
\begin{equation}
\bigg{|}\frac{dF}{dM}\bigg{|} = \frac{d}{dM}\bigg{|}
\int^{\infty}_{0}\!dM^{\prime}
n(M^{\prime})\frac{M^\prime}{\bar\rho}P(M,M^{\prime})\bigg{|}.
\end{equation}
Here $P(M,M^{\prime})$ is the conditional probability of finding a
bound region ($\lambda_{3} > \lambda_{3c}$) 
at filtering mass scale M, provided that it is included
in an isolated bound object ($\lambda_{3} = \lambda_{3c}$) 
with mass $M^{\prime} (> M$).  
The isolated bound objects at a given epoch are those which
have just collapsed. 
Thus, in our formalism, the isolated bound objects 
correspond to the regions with $\lambda_{3} = \lambda_{3c}$ at a 
given filtering mass scale. 

We find that the conditional probability
$P(M,M^{\prime})$ for the case of a sharp k-space filter is given by
(see Appendix B):
\begin{equation}
P(M,M^{\prime}) = 0.08\Theta(M^\prime-M).
\end{equation} 
Equation (29) reveals that $P(\lambda_{3}>0) = 0.08$
results in $P(M,M^{\prime}) = 0.08$ ($M < M^\prime$).  
So, in our formalism the probability of finding 
a bound region ($\lambda_{3} > \lambda_{3c}$) 
of mass scale M included in an isolated bound region 
($\lambda_{3} = \lambda_{3c}$) 
with mass greater M is only $0.08$.  And this is directly related 
to our normalization factor of $1/0.08 = 12.5$. 
Whereas in the PS formalism the probability
$P(M,M^{\prime})$ is $0.5$, which is again 
directly related to the PS normalization factor of $1/0.5 = 2$. 

Now, with equation (28) and (29), we have 
\begin{equation}
\bigg{|}\frac{dF}{dM}\bigg{|} =
0.08\int_{0}^{\infty}dM^\prime n(M^\prime) 
\frac{M^\prime}{\bar{\rho}}
\delta_{D}(M^\prime -M) = 0.08\frac{M}{\bar{\rho}}n(M),
\end{equation} 
where  $\delta_{D}$ stands for the Dirac delta function.
More explicitly, 
\begin{equation}
n(M) = 12.5\frac{\bar{\rho}}{M}\bigg{|}\frac{dF}{dM}\bigg{|}, 
\end{equation}
which is exactly the same formula as equation (2) with the 
normalization factor of $12.5$ included explicitly. 

Thus equation (31) justifies the normalization factor $12.5$ of 
our mass function in the case of a sharp k-space filter.

\section{DISCUSSION AND CONCLUSIONS}

We have derived an {\it analytic} approximation of the mass distribution 
function for clumps.  The underlying dynamics has been described by 
the Zel'dovich approximation which treats the nonspherical 
gravitational collapse.  Similar to Shandarin \& Klypin (1984) 
and Audit et al. (1997), we have assumed that the clumps  
would be formed by the mass elements which have experienced
gravitational collapse along all three directions. 

We have given a somewhat different interpretation to the PS analysis by 
reexpressing the fraction of the volume occupied by the bound regions 
in terms of two probabilities: 
the probability of the Gaussian density distribution 
and the collapse probability of the given dense regions.  
The Zel'dovich approximation has led us to determine a nonspherical 
collapse probability which is different from the PS one, relating 
the density field to the positive lowest eigenvalue of the 
deformation tensor.   

We have shown  that the collapse probability reaches almost unity 
when $\delta > 3\sigma$, which indicates that the highly overdense 
regions will always collapse along all three directions. 
In addition, we have found the density distribution of the regions 
which meet the collapse condition 
based on the Zel'dovich approximation.  
This distribution has shown that the most probable density contrast of 
such regions is around 1.5 at the characteristic mass scale $M_{*}$. 

We have proposed a simple {\it ansatz} in order to treat 
the multistream regions where the Lagrangian dynamics
is not applicable.
This ansatz has enabled us to derive an analytic mass function 
characterized by one free parameter, $\lambda_{3c}$.
The best approximate value of this parameter has been chosen to be 
$\lambda_{3c} \simeq  0.37$. 
We admit that there is no background dynamical theory for
determining directly the value of this free parameter which thus
has to be found phenomenologically. However, the following arguments
may give a hint to understand why this parameter has this value.
A simple extrapolation  of equation (10)
into the multistream regions  
suggests that only the mass elements with
$\lambda_{3} > \lambda_{3c} = 1$ collapse along all three directions 
by the present epoch of $D_{+}=1$. 
However, the collapse along the first two directions increases the
density, which therefore speeds up the collapse along the third directions.
This roughly agrees with the conclusion of Audit et al. (1997).
Using equation (24) from Audit et al. (1997)
with their choice of the
parameters ($\epsilon = 1, \alpha = 0.8, \delta_c = 1.69,
\sigma_c = 0.74$),  one can easily obtain for the collapse epoch
$a_c = 1/(0.8 \lambda_3 + 0.32 \delta)$ that is always earlier
than the prediction of the Zel'dovich approximation ($a_c =1/\lambda_3$)
provided that $\lambda_3 >0$.

For power law spectra, it has been shown that our resulting mass 
function with $\lambda_{3c} \simeq 0.37$ is in good agreement with 
the $\delta_{c}\simeq 1.5$ PS mass function in the high-mass 
section,  but has lower 
peak and predicts more small-mass structures, which are in agreement 
with what has been detected in N-body simulations 
(e.g., \cite{efs-etal88}; \cite{pea-hea90}).
However, it should be noted that the prediction concerning the 
small-mass structures is least reliable not only in any PS-like
approach, but also in our dynamical approach to the mass function
since the validity domain of the Zel'dovich approximation
is limited to the high-mass section as outlined in $\S$ 3.2. 

Like the other PS-like formalisms,
a normalization factor for the mass function has been introduced, 
which in our case is $ 12.5$.   
We have justified the normalization factor 
with a sharp k-space filter by using the Jedamzik integral equation,   
showing that this rather large normalization factor is due 
to the low probability of finding a bound region 
$(\lambda_{3} > \lambda_{3c})$ 
at filtering mass scale $M$ included in an isolated bound region  
($\lambda_{3} = \lambda_{3c}$) with larger mass $M^\prime$.  
But the physical meaning of the sharp k-space filter has yet to
be fully understood.

We postpone the 
numerical testing of our mass function to the following paper. 

\acknowledgments

We are grateful to Lev Kofman, Paolo Catelan, and the  
referee for useful discussions and helpful comments. 
This work has been done under the support of 
NASA grant NAG 5-4039 and EPSCoR 1996 grant.

\newpage

\appendix
\section{APPENDIX}

In 1970 Doroshkevich found the joint probability distribution,  
$p(\lambda_{1},\lambda_{2},\lambda_{3})$ of an 
ordered set of eigenvalues [equation (12) in $\S$ 3.1], 
corresponding to a Gaussian potential. 
In this appendix, we sketch the derivation of
$p(\lambda_{1})$, $p(\lambda_{2})$, and $p(\lambda_{3})$ [equation 
(13), (14) and (15) in $\S$ 3.1], and investigate their statistical 
properties. 

The two point probability distributions, 
$p(\lambda_{1},\lambda_{2}), p(\lambda_{2},\lambda_{3})$ and 
$p(\lambda_{1},\lambda_{3})$ can be easily obtained from the 
direct integration of $p(\lambda_{1},\lambda_{2},\lambda_{3})$ 
such that 
\begin{eqnarray}
p(\lambda_{1},\lambda_{2}) &=& 
\int_{-\infty}^{\lambda_{2}}\!
p(\lambda_{1},\lambda_{2},\lambda_{3})d\lambda_{3} \nonumber \\ 
&=&\frac{1125}{64\sqrt{5}\pi\sigma^4}\Bigg{\{}
(\lambda_{1}-\lambda_{2})(3\lambda_{1}-\lambda_{2})
\exp\bigg{[}-\frac{3\lambda_{1}^2}{\sigma^2} 
+ \frac{3\lambda_{1}\lambda_{2}}{\sigma^2} 
-\frac{9\lambda_{2}^2}{2\sigma^2}\bigg{]} \nonumber \\
&& + \frac{\sqrt{3\pi}\sigma}{12}(\lambda_{1}-\lambda_{2})
\bigg{[}8+\frac{3}{\sigma^2}(3\lambda_{1}-\lambda_{2})
(3\lambda_{2}-\lambda_{1})\bigg{]}
\nonumber \\
&&\times\exp\bigg{[}-\frac{45\lambda_{1}^2}{16\sigma^2} 
+ \frac{15\lambda_{1}\lambda_{2}}{8\sigma^2} 
-\frac{45\lambda_{2}^2}{16\sigma^2}\bigg{]} 
{\rm erfc}\bigg{[}\frac{\sqrt{3}}{4\sigma}
(\lambda_{1}-3\lambda_{2})\bigg{]}\Bigg{\}}
\Theta(\lambda_{1}-\lambda_{2}), \\
\nonumber \\
p(\lambda_{2},\lambda_{3}) &=&
\int^{\infty}_{\lambda_{2}}\!
p(\lambda_{1},\lambda_{2},\lambda_{3})d\lambda_{1} \nonumber \\
&=&\frac{1125}{64\sqrt{5}\pi\sigma^4}\Bigg{\{}
(\lambda_{2}-\lambda_{3})(\lambda_{2}-3\lambda_{3})
\exp\bigg{[}-\frac{3\lambda_{3}^2}{\sigma^2}
+ \frac{3\lambda_{2}\lambda_{3}}{\sigma^2}
-\frac{9\lambda_{2}^2}{2\sigma^2}\bigg{]} \nonumber \\
&& + \frac{\sqrt{3\pi}\sigma}{12}(\lambda_{2}-\lambda_{3})
\bigg{[}8+\frac{3}{\sigma^2}(\lambda_{2}-3\lambda_{3})
(\lambda_{3}-3\lambda_{2})\bigg{]}
\nonumber \\
&&\times\exp\bigg{[}-\frac{45\lambda_{2}^2}{16\sigma^2}
+ \frac{15\lambda_{2}\lambda_{3}}{8\sigma^2}
-\frac{45\lambda_{3}^2}{16\sigma^2}\bigg{]}      
{\rm erfc}\bigg{[}\frac{\sqrt{3}}{4\sigma}
(3\lambda_{2}-\lambda_{3})\bigg{]}\Bigg{\}}
\Theta(\lambda_{2}-\lambda_{3}), \\
\nonumber \\
p(\lambda_{1},\lambda_{3}) &=&
\int_{\lambda_{3}}^{\lambda_{1}}\!
p(\lambda_{1},\lambda_{2},\lambda_{3})d\lambda_{2} \nonumber \\
&=& \frac{1125}{64\sqrt{5}\pi\sigma^4}\Bigg{\{}
(\lambda_{1}-\lambda_{3})(3\lambda_{1}-\lambda_{3})
\exp\bigg{[}-\frac{3\lambda_{1}^2}{\sigma^2}
+ \frac{3\lambda_{1}\lambda_{3}}{\sigma^2}
-\frac{9\lambda_{3}^2}{2\sigma^2}\bigg{]} \nonumber \\
&& + (\lambda_{1}-\lambda_{3})(\lambda_{1}-3\lambda_{3})
\exp\bigg{[}-\frac{3\lambda_{3}^2}{\sigma^2}
+ \frac{3\lambda_{1}\lambda_{3}}{\sigma^2}
-\frac{9\lambda_{1}^2}{2\sigma^2}\bigg{]} \nonumber \\
&& + \frac{\sqrt{3\pi}\sigma}{12}(\lambda_{1}-\lambda_{3})
\bigg{[}\frac{3}{\sigma^2}(3\lambda_{1}-\lambda_{3})
(\lambda_{1}-3\lambda_{3})-8\bigg{]}
\nonumber \\
&&\times\exp\bigg{[}-\frac{45\lambda_{1}^2}{16\sigma^2}
+ \frac{15\lambda_{1}\lambda_{3}}{8\sigma^2}
-\frac{45\lambda_{3}^2}{16\sigma^2}\bigg{]} \nonumber \\
&&\times\bigg{\{}{\rm erfc}\bigg{[}\frac{\sqrt{3}}{4\sigma}
(3\lambda_{1}-\lambda_{3})\bigg{]} - 
{\rm erfc}\bigg{[}\frac{\sqrt{3}}{4\sigma}
(3\lambda_{3}-\lambda_{1})\bigg{]}\bigg{\}}\Bigg{\}}
\Theta(\lambda_{1}-\lambda_{3})
\end{eqnarray}

In order to derive $p(\lambda_{1})$, $p(\lambda_{2})$, and 
$p(\lambda_{3})$, we have to integrate 
the above two point distributions, which  
involve complex error function terms as one can see. 
We find the following recursion formula which is useful 
in integrating such complex terms. 
\begin{eqnarray}
\int\!t^n\exp(-a^2 t^2){\rm erf(bt)}dt & = &
\frac{n-1}{2a^2}\int\!t^{n-2}\exp(-a^2 t^2){\rm erf(bt)}dt \nonumber \\
&&-\frac{1}{2a^2}t^{n-1}\exp(-a^2 t^2){\rm erf(bt)} \nonumber \\
&&+\frac{b}{a^2 \sqrt{\pi}}\int\!t^{n-1}\exp[-(a^2 + b^2)t^2]dt.
\end{eqnarray}
With the above recursion formula, it is straightforward to
derive the individual distributions. The results are 
shown in $\S$ 3.1 [equation (13), (14), and (15)].  

Now the probability that each eigenvalue is positive as well as the mean 
and the variance of each eigenvalue can be computed with the above results:
\begin{eqnarray}
&&P(\lambda_{1} > 0) = \frac{23}{25}, \hspace{1cm}
P(\lambda_{2} > 0) = \frac{1}{2}, \hspace{1cm}
P(\lambda_{3} > 0) = \frac{2}{25}, \\
&&\bar{\lambda_{1}} = \frac{3}{\sqrt{10\pi}}\sigma, \hspace{1cm}
\bar{\lambda_{2}} = 0, \hspace{1cm} 
\bar{\lambda_{3}} = -\frac{3}{\sqrt{10\pi}}\sigma, \\
&&\sigma^{2}_{\lambda_{1}} = \frac{13\pi -27}{30\pi}\sigma^2, 
\hspace{1cm}
\sigma^{2}_{\lambda_{2}} = \frac{2}{15}\sigma^2, \hspace{1cm} 
\sigma^{2}_{\lambda_{3}} = \frac{13\pi -27}{30\pi}\sigma^2, 
\end{eqnarray}
which are all in agreement with Doroshkevich (1970). 

\section{APPENDIX}

In the framework of our formalism, the conditional probability
$P(M,M^{\prime})$ is written as 
\begin{equation}
P(M,M^\prime) = P(\lambda_{3}>\lambda_{3c}|\lambda_{3}^\prime
=\lambda_{3c})
= \frac{P(\lambda_{3}>\lambda_{3c},\lambda_{3}^\prime
=\lambda_{3c})}
{P(\lambda_{3}^\prime=\lambda_{3c})},
\end{equation}
where $\lambda_{3}$ and $\lambda_{3}^\prime$ are the lowest
eigenvalue of the deformation tensor at the same point but at two 
different filtering mass scale $M$ and $M^{\prime}$ respectively. 

In order to derive the probability
$P(\lambda_{3}>\lambda_{3c},
\lambda_{3}^\prime=\lambda_{3c})$,
we start
with the multivariate Gaussian joint probability distribution
(see \cite{do70}; \cite{bar86})
for the six independent elements of the deformation tensor 
at two different mass scales:
\begin{eqnarray}
p_{J}(y_{1},\cdots,y_{6}^{\prime})
dy_{1}\cdots dy_{6}^{\prime}
&=& \frac{\exp(-Q)}{(2\pi)^{6}\sqrt{\det(V)}}
dy_{1}\cdots dy_{6}^{\prime}, \\
Q &=& \frac{1}{2}{\bf y^{t}}\cdot V\cdot{\bf y}
\end{eqnarray}
Here $V$ is the covariance matrix, while  $\{y_{i}\}_{i=1}^{6}$ and  
$\{y_{1}^{\prime}\}_{i=1}^{6}$
are the six independent elements of the deformation tensor 
(defined by equation (11) in $\S$ 3.1; 
$y_{1}\equiv d_{11}$, $y_{2}\equiv d_{22}$, $y_{3}\equiv d_{33}$, 
$y_{4}\equiv d_{12}$, $y_{5}\equiv d_{23}$, $y_{6}\equiv d_{31}$)  
at mass scale $M$ and $M^{\prime}$ respectively.

In the case of a sharp k-space filter, the mutual correlations
between $\{y_{i}\}_{i=1}^{6}$ and $\{y_{1}^{\prime}\}_{i=1}^{6}$ are 
\begin{eqnarray}
&&<y_{i}^2> = \frac{\sigma_{M}^2}{5}, \hspace{1cm}
<y_{i}y_{j}> = \frac{\sigma_{M}^2}{15}, \hspace{1cm}
<y_{i}y_{i}^{\prime}> = \frac{\sigma_{M^\prime}^2}{5},
\\
&&<y_{i}^{\prime 2}> = \frac{\sigma_{M^\prime}^2}{5},
\hspace{1cm}
<y_{i}^\prime y_{j}^\prime> = \frac{\sigma_{M^\prime}^2}{15},
\hspace{1cm}
<y_{i}y_{j}^\prime> = \frac{\sigma_{M^\prime}^2}{15},
\end{eqnarray}
for $i, j(\neq i) = 1,2,3$, and 
\begin{eqnarray}
&&<y_{i}^2> = \frac{\sigma_{M}^2}{15}, \hspace{1cm}
<y_{i}y_{j}> = 0, \hspace{1cm}
<y_{i}y_{i}^{\prime}> = \frac{\sigma_{M^\prime}^2}{15},
\\
&&<y_{i}^{\prime 2}> = \frac{\sigma_{M^\prime}^2}{15},
\hspace{1cm}
<y_{i}^\prime y_{j}^\prime> = 0,
\hspace{1cm}
<y_{i}y_{j}^\prime> = 0,
\end{eqnarray}
for $i, j(\neq i) = 4,5,6$.  
Here $\sigma^2_{M}$ and $\sigma^2_{M^\prime}$
are the mass variance of the density field filtered at the
mass scale $M$ and $M^\prime$ respectively.

Through equation (B2) to (B7), along with    
the similarity transformation
of the deformation tensor into its principal axes, we find
the following joint probability distribution
of the three eigenvalues
$\{\lambda_{i}\}_{i=1}^{3}$, $\{\lambda_{i}^\prime\}_{i=1}^{3}$
of the deformation tensor
at filtering mass scale
$M$, $M^\prime$ $(M < M^\prime)$ respectively.
\begin{eqnarray}
p_{J}(\lambda_{1},\cdots,\lambda_{3}^\prime)
d\lambda_{1}\cdots d\lambda_{3}^\prime
&=& p_{J1}(\Delta_{1},\Delta_{2},\Delta_{3})
d\Delta_{1}d\Delta_{2}d\Delta_{3} \nonumber \\
&&\times p_{J2}
(\lambda_{1}^\prime,\lambda_{2}^\prime,\lambda_{3}^\prime)
d\lambda_{1}^\prime d\lambda_{2}^\prime d\lambda_{3}^\prime
\end{eqnarray}
\begin{eqnarray}
&&p_{J1} =
\frac{5^3\cdot 3^3}{2^4 \pi^3 \sigma_{\Delta}^6 \sqrt{5}}
\exp\bigg{[}-\frac{3I_{1}^2}{\sigma_{\Delta}^2} +
\frac{15I_2}{2\sigma_{\Delta}^2}\bigg{]}
(\Delta_{1}-\Delta_{2})(\Delta_{2}-\Delta_{3})
(\Delta_{3}-\Delta_{1}), \\
&&p_{J2}
= \frac{5^3\cdot 3^3}{2^4 \pi^3 \sigma_{M^\prime}^6 \sqrt{5}}
\exp\bigg{[}-\frac{3I_{1}^{\prime 2}}{\sigma_{M^\prime}^2} +
\frac{15I_{2}^\prime}{2\sigma_{M^\prime}^2}\bigg{]}
(\lambda_{1}^\prime-\lambda_{2}^\prime)
(\lambda_{2}^\prime-\lambda_{3}^\prime)
(\lambda_{3}^\prime-\lambda_{1}^\prime),
\end{eqnarray}
where
\begin{eqnarray}
&&\Delta_{i}\equiv \lambda_{i}-\lambda_{i}^\prime, \hspace{1cm}
\sigma_{\Delta}^2
\equiv \sigma_{M}^2 -\sigma_{M^\prime}^2, \\
&&I_{1} \equiv
\Delta_{1}+\Delta_{2}+\Delta_{3},\hspace{1cm} I_{2}\equiv
\Delta_{1}\Delta_{2}+\Delta_{2}\Delta_{3}+\Delta_{3}\Delta_{1}, \\
&&I_{1}^\prime \equiv \lambda_{1}^\prime +
\lambda_{2}^\prime +\lambda_{3}^\prime, \hspace{1cm}
I_{2}^\prime \equiv
\lambda_{1}^\prime\lambda_{2}^\prime +
\lambda_{2}^\prime\lambda_{3}^\prime +
\lambda_{3}^\prime\lambda_{1}^\prime.
\end{eqnarray}
Note the similarity between $p_{J1}$ and $p_{J2}$.
In fact the above equations hold good only in the case of
a sharp k-space filter.

The integration of $p_{J}$ over
$\lambda_{1},\lambda_{2},\lambda_{1}^\prime$, 
and $\lambda_{2}^\prime$ 
gives us the joint probability density distribution, 
$p(\lambda_{3},\lambda_{3}^\prime)$:
\begin{equation}
p(\lambda_{3},\lambda_{3}^\prime)
d\lambda_{3}d\lambda_{3}^\prime
= p(\Delta_{3})d\Delta_{3}
p(\lambda_{3}^\prime)d\lambda_{3}^\prime.
\end{equation}
Here the probability density distributions of $p(\Delta_{3})$ and
$p(\lambda_{3}^\prime)$ have the same form as $p(\lambda_{3})$ 
[equation (15)] except for the value of the variance.

Finally we derive 
the conditional probability $P(M,M^\prime)$:
\begin{eqnarray}
P(M,M^{\prime}) &=&
 \frac{P(\lambda_{3}>\lambda_{3c},\lambda_{3}^\prime
=\lambda_{3c})}
{P(\lambda_{3}^\prime = \lambda_{3c})}
= \frac{p(\lambda_{3}^\prime=\lambda_{3c})d\lambda_{3}^\prime
\int_{0}^{\infty}\!d\Delta_{3c}p(\Delta_{3c})}
{p(\lambda_{3}^\prime =\lambda_{3c})d\lambda_{3}^\prime} \nonumber \\
&=& \int_{0}^{\infty}\! 
d\Delta_{3c}p(\Delta_{3c})
= \int_{0}^{\infty}\! d\lambda_{3}p(\lambda_{3})
= 0.08\Theta(M^\prime-M), 
\end{eqnarray}
where $\Delta_{3c}$ is $\lambda_{3}-\lambda_{3c}$.

\begin{figure}[tb]
\psfig{figure=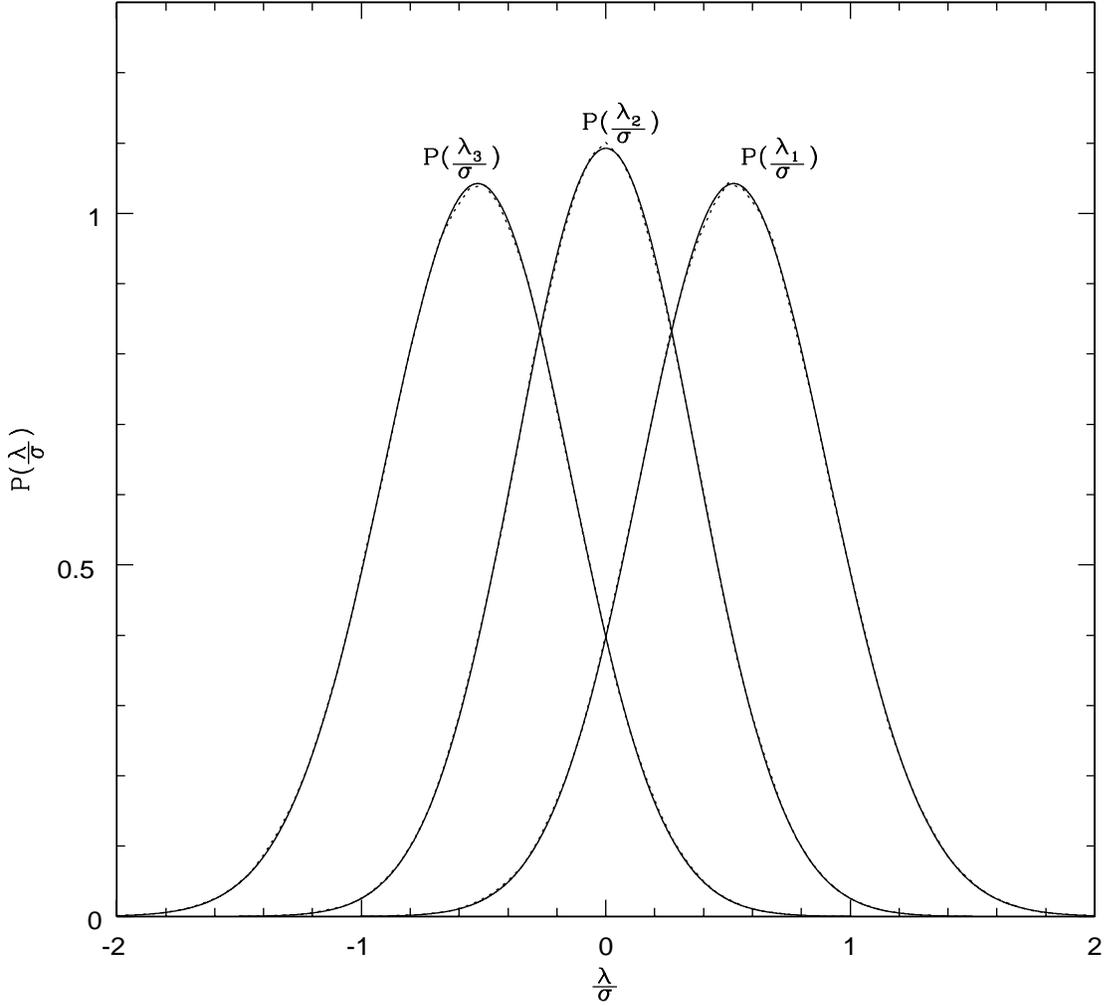,height=14.5cm,width=15.5cm}
\caption{The individual probability distributions of 
three eigenvalues of the deformation tensor : The solid lines
shows the analytic results obtained in this paper 
for the rescaled variable $\lambda/\sigma$. 
The numerical results from the Monte Carlo simulation 
are also plotted as the dotted lines.}
\end{figure}
\begin{figure}[tb]
\psfig{figure=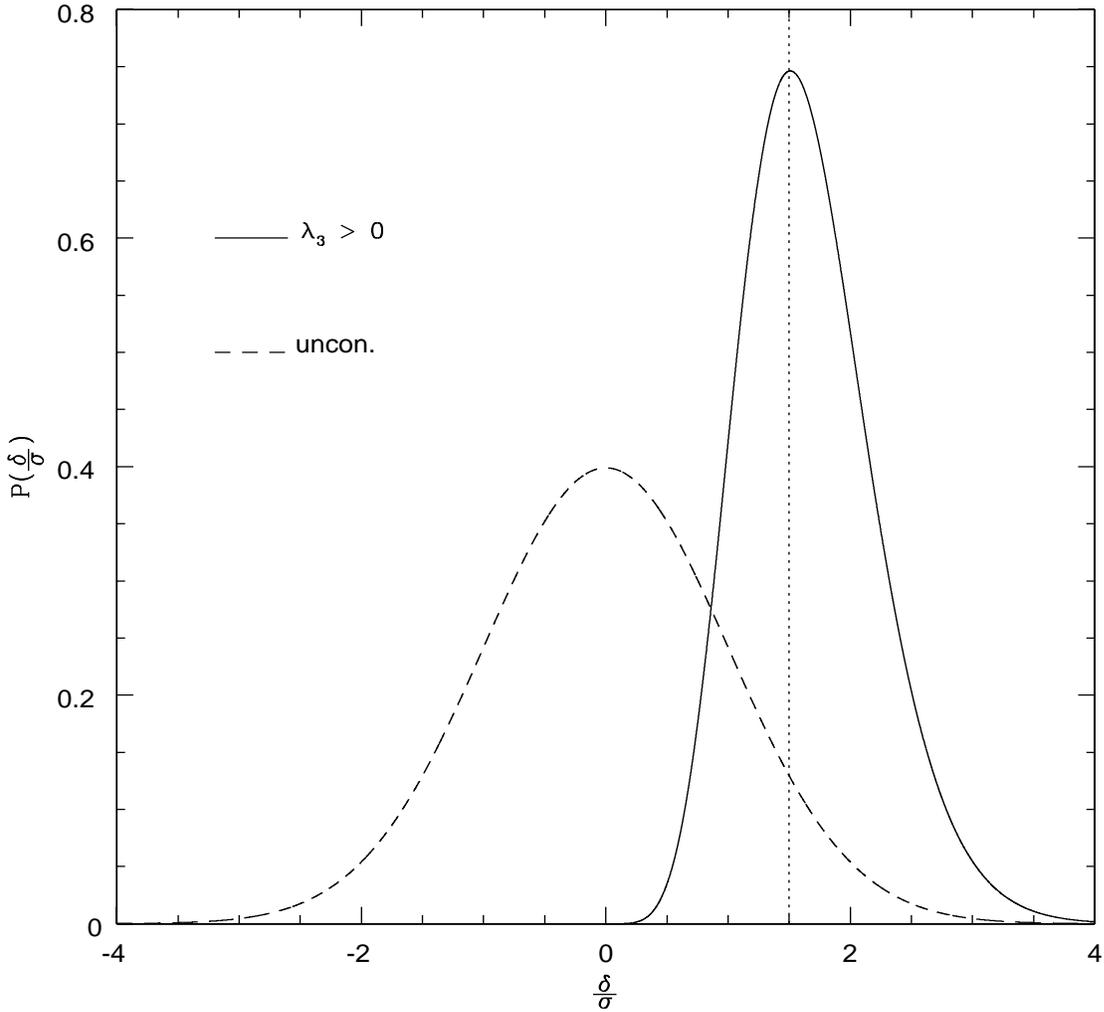,height=14.5cm,width=15.5cm}
\caption{The probability distribution of the rescaled density field,  
$(\delta/\sigma)$. 
The solid line represents the rescaled density  
distribution under the condition of $\lambda_{3} > 0$,  
while the dashed line shows the unconditional Gaussian 
distribution.  The vertical dotted line indicates the position of 
$\delta/\sigma =1.5$}
\end{figure}
\begin{figure}[tb]
\psfig{figure=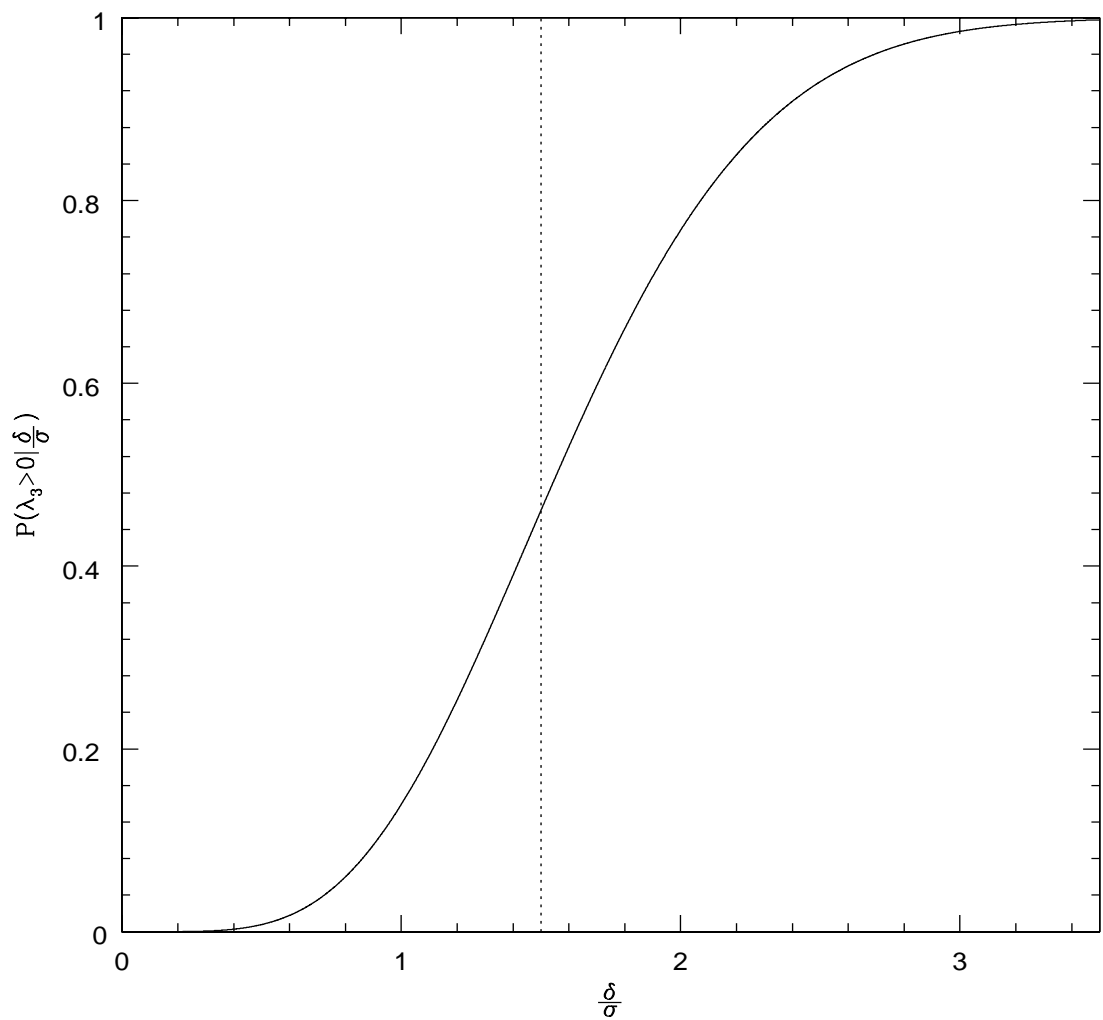,height=14.5cm,width=15.5cm}
\caption{The conditional probability of $\lambda_{3} > 0$ as
a function of the rescaled density $\delta/\sigma$.  
The vertical dotted line indicates the position $\delta/\sigma =1.5$.}
\end{figure}
\begin{figure}[tb]
\psfig{figure=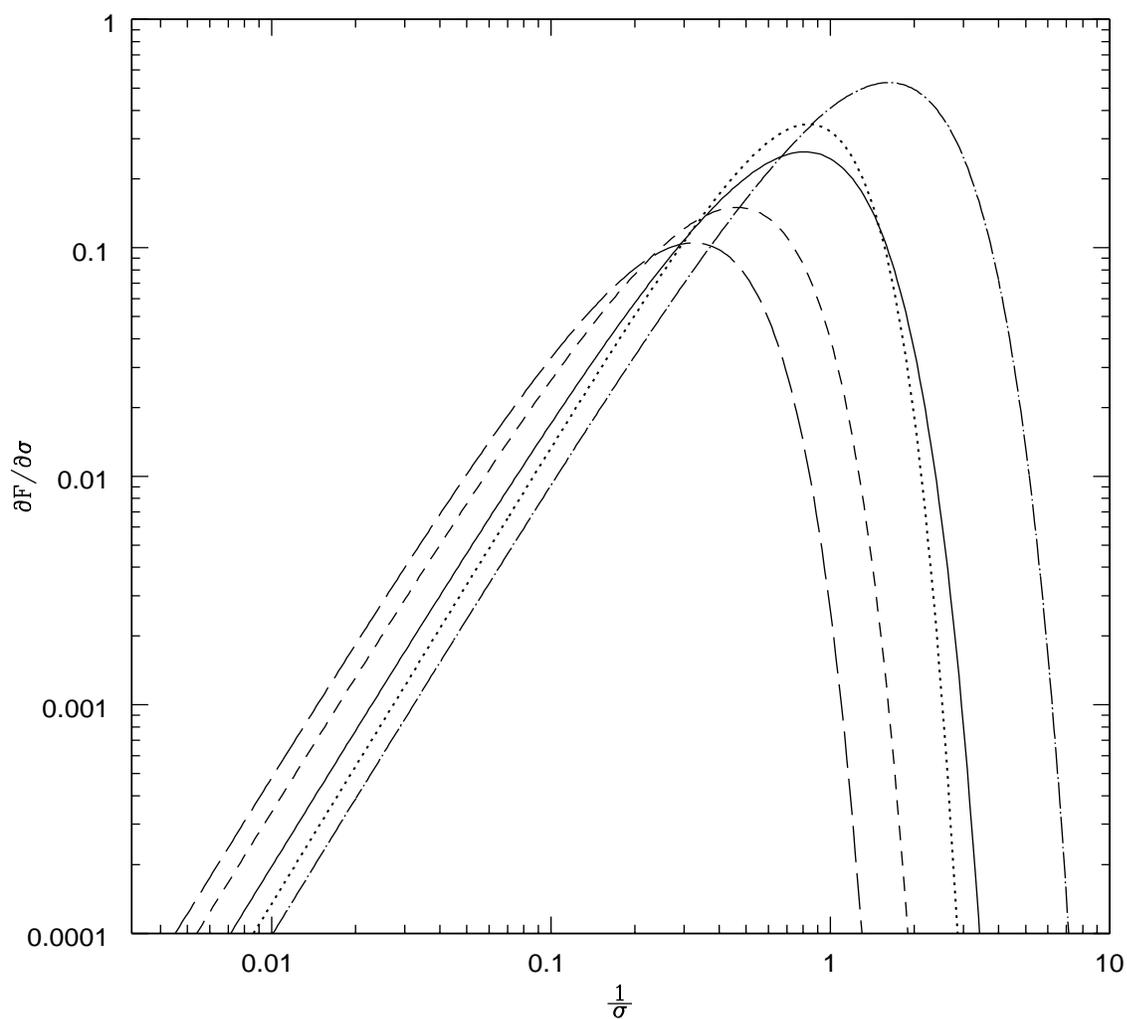,height=14.5cm,width=15.5cm}
\caption{The differential volume fraction for $\lambda_{3c}$
= $0.1,0.4,0.7$ and $1.0$ (dot-dashed, solid, dashed and long-dashed
lines).  The dotted line is the standard 
$(\delta_{c}\simeq 1.69)$ PS differential volume fraction.}
\end{figure}
\begin{figure}[tb]
\psfig{figure=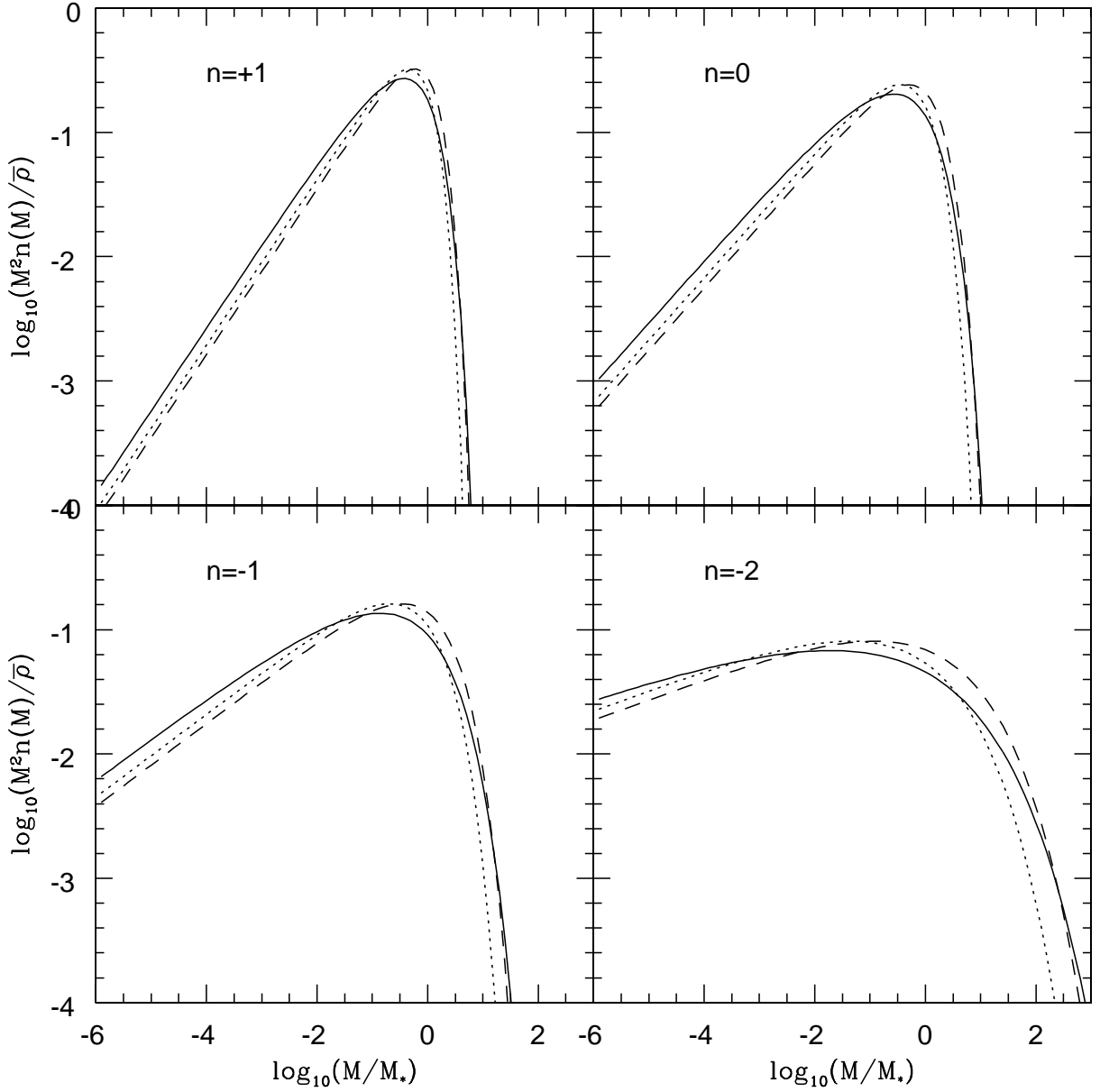,height=17.5cm,width=17.5cm}
\caption{The mass function for the power index $n=+1,0,-1$ and $-2$.
The solid line shows our analytic results with $\lambda_{3c} = 0.37$,
while the  dashed and dotted lines represent the PS mass function
with $\delta_{c} = 1.5$ and $1.69$ respectively.}
\end{figure}
\end{document}